\documentclass[conference]{IEEEtran}
\usepackage{cite}

\ifCLASSINFOpdf
\else
\fi
\usepackage[cmex10]{amsmath}
 \usepackage{stfloats}

\usepackage[dvips]{graphicx}

\usepackage{amssymb}

\newtheorem{lemma}{Lemma}
\newtheorem{theorem}{Theorem}

\begin{document}
%
\title{Average Secrecy Capacity of Free-Space Optical Communication Systems with On-Off Keying  Modulation and Threshold Detection}

\author{\IEEEauthorblockN{Jinxiao~Zhu\IEEEauthorrefmark{1},
Yin~Chen\IEEEauthorrefmark{2}, and
Masahide~Sasaki\IEEEauthorrefmark{1}}
\IEEEauthorblockA{\IEEEauthorrefmark{1} Quantum ICT Laboratory, Advanced ICT Research Institute\\
 National Institute of Information and Communications Technology (NICT), Tokyo 184-8795, Japan }
\IEEEauthorblockA{\IEEEauthorrefmark{2}Graduate School of Media and Governance\\
Shonan Fujisawa Campus, Keio University, Fujisawa-shi, Kanagawa 252-0882, Japan\\ 
Email: \{zhu, psasaki\}@nict.go.jp; yin@ht.sfc.keio.ac.jp}}

\markboth{Journal of \LaTeX\ Class Files,~Vol.~6, No.~1, January~2007}%
{Shell \MakeLowercase{\textit{et al.}}: Bare Demo of IEEEtran.cls for Journals}
%



\maketitle
\thispagestyle{empty}

\begin{abstract}
This paper considers secure communication over free-space optical (FSO) links suffering from turbulence-induced fading. 
In particular, 
we study the average secrecy capacity of applying physical layer security to achieve security in an FSO communication system with binary-level on-off keying intensity modulation and threshold detection,
a widely used system in FSO communication.
We first define the instantaneous secrecy capacity for our system and then provide a lower bound of it, 
which is achieved based on a simple design principle for practical applications and yields a simpler equation.
We next derive a general expression of the average secrecy capacity and the corresponding lower bound based on bit error probabilities.
Finally, numerical analysis is conducted for the special case of the correlated log-normal fading model
to show the impacts of fading and channel correlation on the average secrecy capacity.

\end{abstract}


%
\IEEEpeerreviewmaketitle

\section{Introduction}
%
%
%
%
Laser-based  free-space optical (FSO) communication, 
a communication technology that uses high-directional laser beam to transmit data in free space, has attracted considerable attention recently due to its merits, like high-data-rate and cost-effective \cite{Khalighi2014Survey}.
Although it has high directionality, 
FSO communication still suffers from the security issue because of the openness of wireless medium, 
especially when the main lobe of the laser beam is considerably wider than the receiver size \cite{Wang2014Enhancing}. 
Thus, solving the security issue is critical to support its application in some mission-critical applications, like satellite communication and military communication.

Recently, there has been a great interest in applying physical layer security to achieve information security in wireless communication systems, 
since it enables tractable performance analysis and also enables positive secrecy data rate even when the channel to the target receiver is in worse condition than that to an eavesdropper \cite{Leung-Yan-Cheong1978ITIT, Barros2006Secrecy}.
In FSO communication, channel condition depends strongly on the atmospheric environment between the transmitter and the receiver, 
which keeps varying over time and space randomly \cite{Andrews2005Laser, Nistazakis2009Average}.
Therefore, physical layer security schemes which adapt to the change of the channel condition are really important in FSO communications.





Secrecy capacity, defined as the maximum information rate at which the source can transmit to the receiver without the eavesdropper being able to acquire any information, is a key metric to assess the security level of an FSO system based on physical layer security.
By now, although extensive theoretical and experimental performance analysis has been conducted for FSO communications, only few studies on secrecy capacity of FSO systems have been done.
\cite{Lopez-Martinez2015Physical} characterized the probability of positive secrecy capacity in the presence of atmospheric turbulence-induced random fading for two eavesdropping scenarios.
In \cite{Wang2014Enhancing}, Wang et al. studied the capacity of secret key generation for air-to-ground FSO communication.
And the secrecy capacity for indoor visible light communication without fading phenomenon based on multiple-input and single-output (MISO) Gaussian model was analyzed in \cite{Mostafa2014Physical}. 
More recently, numerical study on the secrecy capacity has been conducted in \cite{Endo2015Numerical} for FSO channels based on the on-off keying (OOK) modulation and photon counting demodulation without considering the random fading effect.


In this paper, we study the secrecy capacity of laser-based FSO communication systems where OOK intensity modulation with direct detection (IM/DD) is employed.  
Notice that, due to their simplicity and low cost, IM/DD systems are widely used in FSO communications \cite{Khalighi2014Survey}.
Our contributions in this paper are summarized as follows.
\begin{itemize}
  \item We define the instantaneous secrecy capacity of the FSO system with binary-level OOK intensity modulation and threshold detection under a given pair of fading states.
	\item We derive a lower bound of the instantaneous secrecy capacity of the fading FSO system
	based on bit error probabilities, and prove that this lower bound 
	is just the maximum instantaneous secrecy rate when the detection threshold at the eavesdropper is halfway between the expected outputs of the binary input symbol `0' and `1'. 
	\item With the help of the above result, we derive a general expression of the average secrecy capacity of the fading FSO system and also its lower bound.
	\item By specializing the average secrecy capacity to the correlated log-normal fading model, we provide numerical analysis to explore the impacts of fading, channel correlation and the atmospheric turbulence (fading) strength on the secrecy capacity, respectively.
\end{itemize}

The remainder of this paper is organized as follows. 
Section \ref{sec_sysModelMetric} presents the system model and performance metrics for the FSO system.
In Section \ref{sec_secCapacity}, we provide a lower bound of the instantaneous secrecy capacity 
and derive a general formula of the average secrecy capacity for the fading FSO system. 
In Section \ref{sec_dicussion}, we specialize the above general formula to the correlated log-normal fading model
and conduct numerical analysis.
Finally, we conclude the paper in Section \ref{sec_conclusion}




\section{System Model and Metrics} \label{sec_sysModelMetric}
In this section, we introduce the system model and performance metrics. 
The notation and symbols used throughout the paper are summarized in Table \ref{table_notation}.
Note that $C$ denotes the capacity of a general channel while $C_b$ is the capacity of the channel to Bob, 
and the similar rule will be used in other cases when there is no fear of confusion.
By default, $\log (\cdot )$ refers to the natural logarithm.

\subsection{System model} \label{sec_sysModel}
We consider a laser-based FSO system in which a transmitter (named Alice) tries to transmit confidential messages to a receiver (named Bob) over a wireless optical link in the presence of an eavesdropper (named Eve).
In the FSO system， we consider that the binary-level on-off keying modulation and threshold detection are adopted \cite{Hranilovic2006Wireless, Andrews2001Laser}.

%
%

The channel model of the FSO communication system is given by the expression 
\vspace*{-5  pt}
\begin{equation}  \label{eq_sysModel1}
   Y_u =  \eta I_{tx} e^{- \delta d_u} H_u X + Z_u,
		\vspace*{-5  pt}
\end{equation}
where $u \in \{b,e\}$ denotes the user Bob or Eve, $Y_u$ is the electric current of the signal received at $u$, 
$\eta$ is photo-diode responsivity, 
$I_{tx}$ denotes the radiant emittance of the laser beam at transmitter, 
$e^{- \delta d_u}$ represents the intensity attenuation loss in free space (mainly due to absorption and scattering), $\delta$ being an attenuation loss coefficient and $d_u$ the distance from the transmitter to $u$ \cite{Ghassemlooy2012Optical}, 
$H_u$ is a random variable (RV) that models the normalized optical irradiance fluctuation of the channel to $u$ (caused by refractive-index variations), 
$X \in \{ 0,1 \}$ is the binary input symbol, 
and $Z_u \sim  \mathcal{N} (0, \sigma_u^2)$ models the combined effects of the thermal noise and the background noise at the receiver \cite{Wang2014Enhancing}. 
The system model in (\ref{eq_sysModel1}) can be simplified as
%
\vspace*{-5  pt}
\begin{equation}  \label{eq_sysModel2}
   Y_u = S_{u} H_u X + Z_u,
	\vspace*{-4  pt}
\end{equation}
where $S_{u} = \eta I_{tx} e^{- \delta d_u} $ represents the electric current received at $u$ for the on-state input (symbol `1') in the absence of atmospheric turbulence-induced irradiance fluctuation.
The input probability of $X$ is given by $P_X(1) = q$ and $P_X(0) = 1-q$ with $q \in [0,1]$.
The detection threshold at Bob and Eve are denoted by $\tau _b$ and $\tau _e$, respectively.

In this model, the maximum power constraint is imposed for the transmission of each symbol, 
that is, the radiant emittance for symbol `1' is just the maximum value $I_{tx}$ and that for symbol `0' is zero. 
We assume $\sigma_b = \sigma_e$, in order to focus on exploring the impact of fading on the secrecy capacity.


The codeword corresponding to a confidential message is assumed to be transmitted over one coherence interval.
It is also assumed that both Bob's and Eve's channels are under quasi-static fading, that is, 
$H_b$ and $H_e$, albeit random, are constants during the transmission of an entire codeword (i.e., a coherence interval) and, moreover, vary independently from codeword to codeword.
Bob and Eve are assumed to know their own channel states $H_b$ and $H_e$ respectively, 
and Alice is assumed to know both channel states before the transmission of the codeword.   
This assumption is made to analyze the average secrecy capacity and it corresponds to the situation, for example, where Eve is another active user in the wireless TDMA network \cite{Bloch2008ITIT}.

%

\begin{table}[!t]
\renewcommand{\arraystretch}{1.3}
\caption{Summary of Notations}
\label{table_notation}
\centering
\begin{tabular}{|l|l|}
\hline
Symbol & Meaning\\
\hline
$u $ & $u \in \{b,e\}$ denotes the user Bob or Eve\\
$Y_u$ & Signal current received at $u$ \\
$X$ & $X \in \{ 0,1 \}$ denotes the input symbol \\
$Z_u$ & $Z_u \sim  \mathcal{N} (0, \sigma_u^2)$ denotes the Gaussian distributed noise at $u$ \\
      & with zero-mean and variance $ \sigma_u^2$\\
$\eta$ & Photo-diode responsivity \\
$I_{tx}$ & Radiant emittance of the laser beam at the transmitter Alice\\
$\delta $ & Intensity attenuation loss constant in free space \\
$d_u$ & Distance from the transmitter to $u$\\
$\tau _u$ & User $u$'s detection threshold \\
$H_u$ & A random variable (RV) that models the mean-normalized  \\
      & irradiance fading of the channel to $u$\\
$S_{u} $ & $S_{u} = \eta I_{tx} e^{- \delta d_u}$ denotes the signal current received at $u$  \\
      & for symbol `1' without turbulence-induced fading\\
$\gamma_u$ & $\gamma_u=  \frac{S_u}{\sigma_ u}$ denotes the average electrical SNR at $u$\\
$\overline{y}_u$ &  Mean value of the output signal current under the state $h_u$ \\
$T_u$ & $T_u \sim  \mathcal{N} (\mu_{T_u}, \sigma _{T_u} ^2)$ is a Gaussian distributed RV \\
      & and related to $H_u$ as $H_u = \exp (T_u)$ \\
$q$ & Transmission probability of input `1', i.e.,  $P_X(1) = q$\\
$\rho$ & Correlation coefficient between $T_b$ and $T_e$ \\
$\rho_H$ & Correlation coefficient between $H_b$ and $H_e$\\
$\varepsilon^b$, $\varepsilon^e$ & Error probabilities at Bob and Eve, resp.\\
$C_s$ & Secrecy capacity\\
$C_u$ & Capacity of the channel from Alice to $u$\\
$\mathbb{P}(\cdot)$ & Probability operator \\
$\mathbb{E}(\cdot)$ & Expectation operator \\
\hline
\end{tabular}
\end{table}

\subsection{Performance metrics} \label{sec_metrics}
In a communication system, secrecy capacity is a key metric to measure the tightest upper bound on the average amount of information that can be reliably transmitted over the channel to the target receiver  
while keeping the eavesdropper totally ignorant about the information. 
In this paper, we will use the secrecy capacity to measure the performance of applying the physical layer security scheme in the considered system.
Before defining the secrecy capacity, we will introduce the definition of capacity in our model in the hope of making the  
definition of secrecy capacity easier to understand. 

Usually, capacity is a function of the matrix of channel transition probabilities in a discrete memoryless channel \cite{Gallager1968Information}, for instance, a function of the detection threshold $\tau$ under a given realization of fading $h$ in our system, i.e., $C (\tau; h) = \max_{q } I(X;Y)$,
where $I(\cdot)$ refers to the mutual information \cite{Gallager1968Information}.
This is because for the considered FSO channel under a gvien fading state, the channel can be viewed as a binary asymmetric channel with crossover probabilities for `0' and `1' determined by the detection threshold.
However, in the considered system, we can adjust the detection threshold $\tau$ and also the input probability parameter $q$, thus the instantaneous capacity is given by $C (h) = \max_{q, \tau } I(X;Y)$ correspondingly\footnote{In the context of fading channel, the capacity for a particular fading state is called instantaneous capacity and the expected value over the domain of fading state is called average capacity. }.  
Since Alice cannot control and also doesn't know Eve's detection threshold, the secrecy capacity should be achieved by assuming that Eve adopts its optimal detection threshold for any $q$.
Since $\sigma_b = \sigma_e$ is assumed, we are able to construct a wire-tap channel such that Bob's channel is more capable than Eve's channel
 when $h_b > h_e$ \cite{Csiszar2011Information}.
\emph{Hence, the instantaneous secrecy capacity of the considered system for fading states $h_b$ and $h_e$ is defined by}
\begin{equation}  \label{eq_secCapDef}
   C_{s} (h_b, h_e) = \max_{q, \tau_b } \{ I(X;Y_b) - \max_{\tau_e } I(X;Y_e) \}.
\end{equation}
Meanwhile, when Alice and Eve's detection thresholds are given, the maximum instantaneous secrecy rate is defined by
\begin{equation}  \label{eq_SecRateDef}
   C_{s}(\tau_b, \tau_e; h_b, h_e) = \max_{q} \{ I(X;Y_b) -  I(X;Y_e) \}.
\end{equation}


\section{Secrecy Capacity Analysis}  \label{sec_secCapacity}
In this section, we will derive the average secrecy capacity and its lower bound based on bit error probabilities.

\subsection{Bit error probability}
In binary-level OOK, each bit symbol is transmitted by pulsing the light source either on (for input symbol `1') or off (for `0') during each bit time.
In the corresponding threshold detection, we say a `1' is received when the output of the detector exceeds a threshold value, or else a `0' is received during each bit time \cite{Andrews2001Laser}.
Because of noise, errors may be made at Bob or Eve in determining the actual symbols transmitted.
For instance, a `0' might be mistaken for a `1', which is denoted by $\mathbb{P} (1|0)$.
Also a `1' might be mistaken for a `0', which is denoted by $\mathbb{P} (0|1)$.
In the case a `0' is transmitted, the received signal is observed to be independent of irradiance fading state, 
which is given by $Y_u \sim  \mathcal{N} (0, \sigma_u^2)$ with $u \in \{b,e\}$.
On the other hand, when a `1' is transmitted, 
the received signal under a given realization $h_u$ of fading state $H_u$ is given by  $Y_u \sim  \mathcal{N} (\overline{y}_u, \sigma_u^2)$, where $\overline{y}_u = S_u h_u$ denotes the mean value of output signal current under the state $h_u$.

For a given detection threshold $\tau _u$ at user $u$, the bit error probabilities of $u$'s channel can be derived by
\begin{IEEEeqnarray}{rCl}
   \mathbb{P}_u (1|0) &=&  \frac{1}{\sqrt{2 \pi} \sigma_u} \int_{\tau _u}^{+ \infty } \mathrm{e}^{- \frac{y^2}{2 \sigma_ u^2}} \mathrm{d} y    \nonumber \\
    &=& \frac{1}{2} \mathrm{erfc} \left( \frac{\tau _u}{ \sqrt{2 } \sigma_u} \right )  　 \label{eq_BERBob0_tmp}
\end{IEEEeqnarray}
and
\begin{IEEEeqnarray}{rCl}
    \mathbb{P}_u (0|1) &=&  \frac{1}{\sqrt{2 \pi} \sigma_u} \int_{-\infty }^{\tau _u} \mathrm{e}^{- \frac{ (y - \overline{y}_u) ^2}{2 \sigma_ u^2}} \mathrm{d} y    \nonumber \\
    &=& \frac{1}{2} \mathrm{erfc} \left( \frac{\overline{y}_u   -  \tau _u }{ \sqrt{2 } \sigma_ u} \right )    \label{eq_BERBob1_tmp}
\end{IEEEeqnarray}
under the state $h_u$.
Here, $\mathrm{erfc} (x) =  \frac{2}{\sqrt{\pi}}  \int_{x}^{\infty } \mathrm{e}^{- t^2} \mathrm{d} t$ is the complementary error function. 

A special scenario of the detection thresholds that 
has been widely used in the literature, see \cite{Andrews2001Laser, Wang2012Performance}, are
{\setlength\arraycolsep{0.1em}
\begin{eqnarray}
\tau _b &=& \frac{\mathbb{E}(Y_b|X=0,h_b) + \mathbb{E}(Y_b|X=1,h_b) }{2},    \label{eq_threshBob}  \\  
\tau _e &=& \frac{\mathbb{E}(Y_e|X=0,h_e) + \mathbb{E}(Y_e|X=1,h_e) }{2}  \label{eq_threshEve}
\end{eqnarray}
}%
where $\mathbb{E}(Y_b|X=1,h_b)$ refers to the Bob's expected signal amplitude for bit `1' conditioned on a particular fading state $h_b$, and the other similar notations can be explained likewise. 
In our channel model, we have $\mathbb{E}(Y_u|X=1,h_u) = S_u h_u$  and $\mathbb{E}(Y_u|X=0,h_u) = 0$ with $u \in \{b,e\}$.
Then, the bit error probabilities under this special scenario are
\begin{equation}  \label{eq_BERBob01}
   \mathbb{P}_b (1|0) = \mathbb{P}_b (0|1)  =  \frac{1}{2} \mathrm{erfc} \left( \frac{S_b h_b }{2 \sqrt{2 } \sigma_ b} \right )
\end{equation}
and 
\begin{equation}  \label{eq_BEREve01}
   \mathbb{P}_e (1|0) = \mathbb{P}_e (0|1)  =  \frac{1}{2} \mathrm{erfc} \left( \frac{S_e h_e }{2 \sqrt{2 } \sigma_ e} \right ).
\end{equation}
For simplicity, let $ \mathbb{P}_b (1|0) = \mathbb{P}_b (0|1) = \varepsilon^b$
and $ \mathbb{P}_e (1|0) = \mathbb{P}_e (0|1) = \varepsilon^e$. 
This shows that the ``halfway'' settings of the threshold value in (\ref{eq_threshBob}) and (\ref{eq_threshEve}) lead to 
binary symmetric channels with crossover probabilities $\varepsilon^b$ for Bob's channel and $\varepsilon^e$ for Eve's channel.


\subsection{Lower bound of instantaneous secrecy capacity} \label{sec_preliminaries}
Finding a closed-form result of $C_s$ is often a tedious if not impossible task as the maximization operation relies on  several parameters 
like the channel fading states $H_b$ and $H_e$, noises $Z_b$ and $Z_e$, and detection thresholds $\tau_b$ and $\tau_e$.
We now derive a lower bound of the instantaneous secrecy capacity under a given pair of states $h_b$ and $h_e$.
\begin{IEEEeqnarray}{rCl}
    \lefteqn{
      C_{s} (h_b, h_e)
    } \nonumber \\ \quad
   &\geq&  \left [ \max_{ q = \frac{1}{2}, \tau_b } \{ I (X;Y_b) - \max_{\tau_e } I(X;Y_e) \} \right ]^+ \nonumber \\
    &=& \left [  \max_{\tau_b } I \left( X;Y_b | q = \frac{1}{2} \right)　- \max_{\tau_e } I \left(X;Y_e | q = \frac{1}{2} \right) \right ]^+  \nonumber   \\
		&=&  \underline{C_{s}(h_b, h_e)}  \nonumber
\end{IEEEeqnarray}
where $[x]^+ = \max \{0, x \}$, $I \left( X;Y| q \right)$ is the mutual information between $X$ and $Y$ conditioned on $q$.
We use $\underline{C_{s} (h_b, h_e)}$ to denote this lower bound.
In order to calculate $\underline{C_{s} (h_b, h_e)}$, we have the following lemma.

\begin{lemma} \label{lemma_optimumThresh}
When $q = 1/2$,
the eavesdropper channel's maximum information rate under the fading state $h_e$, $ \max_{\tau_e } I \left(X;Y_e | q = \frac{1}{2} \right) $, is achieved by the ``halfway'' setting of the threshold value in (\ref{eq_threshEve}).
\end{lemma}
\begin{IEEEproof}
See Appendix $\ref{sec_proofLemma1}$.
\end{IEEEproof}
%

From Lemma \ref{lemma_optimumThresh}, we have
\begin{equation} \label{eq_capacityLB}
    \max_{\tau_e } I \left(X;Y_e | q = \frac{1}{2} \right) =  1 -  \mathcal{H}(\varepsilon ^e).
\end{equation}
Although Lemma \ref{lemma_optimumThresh} is built for Eve's channel, it also applies to Bob's channel.
Therefore, we can derive the lower bound of secrecy capacity for any $h_b$ and $h_e$ as 
\begin{equation} \label{eq_secCapacityLB}
    \underline{C_{s}(h_b, h_e)} =  \left [  \mathcal{H}(\varepsilon ^e) -  \mathcal{H}(\varepsilon ^b) \right]^+,
\end{equation}
where $\mathcal{H}(\cdot)$ is the binary entropy function given by
\begin{equation} \label{eq_binaryEntropy}
    \mathcal{H}(p) = -p \log _2 p - (1-p) \log _2 (1-p).
\end{equation}

\begin{lemma} \label{lemma_optimumX}
When  $\tau _b$ and $\tau _e$ adopt the ``halfway'' settings given in (\ref{eq_threshBob}) and (\ref{eq_threshEve}),
the maximum instantaneous secrecy rate is given by
\begin{equation} \label{eq_secCapacity_threshold}
    C_{s} \left(\tau _b =  \frac{\overline{y}_b}{2}, \tau_e = \frac{\overline{y}_e}{2}; h_b, h_e \right)  = \left [  \mathcal{H}(\varepsilon ^e) -  \mathcal{H}(\varepsilon ^b) \right]^+.
\end{equation}
and it is achieved when the input probability $q = 1/2$.
\end{lemma}
\begin{IEEEproof}
When $\tau _b =  \overline{y}_b/ 2$ and $ \tau_e = \overline{y}_e / 2 $, 
the channel can be viewed as a binary symmetric broadcast channel (BSBC).
The secrecy capacity of BSBC is given by \cite[Lemma~1]{maurer1993secret}
\begin{equation*} 
    C_{s} \left(\tau _b =  \frac{\overline{y}_b}{2}, \tau_e = \frac{\overline{y}_e}{2}; h_b, h_e \right) = \left [       \mathcal{H}(\varepsilon ^e) -  \mathcal{H}(\varepsilon ^b) \right]^+.
\end{equation*}
\end{IEEEproof}

From Lemma \ref{lemma_optimumThresh} and Lemma \ref{lemma_optimumX}, we observe that the maximum instantaneous secrecy rate 
under the ``halfway'' settings of the threshold value in (\ref{eq_threshBob}) and (\ref{eq_threshEve}) 
equals to the lower bound of instantaneous secrecy capacity, that is 
\begin{equation}  \label{eq_SecRate_special}
   C_{s} \left(\tau _b =  \frac{\overline{y}_b}{2}, \tau_e = \frac{\overline{y}_e}{2}; h_b, h_e \right) = \underline{C_{s}(h_b, h_e)}.
\end{equation}






\subsection{Average secrecy capacity}  \label{subsec_secCap}

Like the calculation of average secrecy capacity of fading wiretap channel in wireless radio-frequency (RF) communication \cite{Gopala2008ITIT}, 
the average secrecy capacity for the concerned FSO model is similarly calculated by first deriving the instantaneous secrecy capacity for a given pair of states $h_b$ and $h_e$ and then integrating them over the domain of $(H_b, H_e)$.
Thus, we establish the following theorem on the average secrecy capacity of the concerned FSO system and a lower bound to simplify its calculation.
\begin{theorem} \label{lemma_secCapacity}
For the concerned binary-level OOK modulation and threshold detection, 
the average secrecy capacity of the FSO channel is given by
\begin{IEEEeqnarray}{rCl}\label{eq_avgSecCapacity}
   C_s		&=&   \!  \int_{0}^{\infty } \!\!\!\! \int_{0}^{\infty } \!\!\!  C_{s} (h_b, h_e)  f_{H_b, H_e} (h_b, h_e)  \mathrm{d} h_b \mathrm{d} h_e, \ \ \ \ \ 
\end{IEEEeqnarray}
where $C_{s} (h_b, h_e)$ is given in (\ref{eq_secCapDef}) 
and $f_{H_b, H_e} (h_b, h_e)$ is the joint probability density function (PDF) of the RVs $H_b$ and $H_e$.
Furthermore, the average secrecy capacity is lower bounded by 
\begin{IEEEeqnarray}{rCl}\label{eq_avgSecCapacityLB}
   \underline{C_s} 		&=&   \!  \int_{0}^{\infty } \!\!\!\! \int_{0}^{\infty } \!\!\!  \left[ \mathcal{H} \left(\varepsilon^e \right ) - \mathcal{H} \left(\varepsilon^b \right ) \right ]^+  f_{H_b, H_e} (h_b, h_e)  \mathrm{d} h_b \mathrm{d} h_e, \ \ \ \ \ 
\end{IEEEeqnarray}
where $\mathcal{H}(\cdot)$ is given in (\ref{eq_binaryEntropy}), 
and $ \varepsilon^b$ and $\varepsilon^e$ are given in ($\ref{eq_BERBob01}$) and ($\ref{eq_BEREve01}$), respectively.
\end{theorem}

\begin{IEEEproof}
The proof for the average secrecy capacity is quite simple. 
For a given pair $h_b$ and $h_e$, the instantaneous secrecy capacity is given by the definition (\ref{eq_secCapDef}).
Since $H_b$ and $H_e$ vary independently from codeword to codeword, the average secrecy capacity can be derived by integrating the instantaneous secrecy capacity over the regions of $H_b$ and $H_e$.  

We now prove the lower bound.
The lower bound of the instantaneous secrecy capacity under a given pair of states $h_b$ and $h_e$ has been derived in (\ref{eq_secCapacityLB}).
Hence, the lower bound of the fading FSO channel can be derived by integrating this bound over the regions of $H_b$ and $H_e$
\begin{equation}  \label{eq_secCapacity1}
   \underline{C_s} =  \int \int \underline{C_{s}(h_b, h_e)}  f_{H_b, H_e} (h_b, h_e) \mathrm{d} h_b \mathrm{d} h_e.
\end{equation}
Thus, the result in (\ref{eq_avgSecCapacityLB}) is proved.
\end{IEEEproof}

It is worth noting that $\underline{C_s}$ is estimated to be tight since the ``halfway'' setting of the detection threshold 
yields a minimum error probability\footnote{It is observed that the overall error probability $ \mathbb{P}(1|0) + \mathbb{P} (0|1) $  is minimized at the detection threshold under the ``halfway'' setting.}.
Moreover, it is notable that the lower bound, which greatly simplified our secrecy capacity formula, 
 can be achieved based on a simple design principle in practical applications.


\newcounter{tempequationcounter1}
\setcounter{tempequationcounter1} {\value{equation}}
\begin{figure*}[!bh]
  \normalsize
	\hrulefill
\vspace*{1  pt}
	\setcounter{equation} {20}
\begin{IEEEeqnarray}{rCl}  
    \IEEEeqnarraymulticol{3}{l}{
     f_{H_b, H_e} (h_b, h_e)    =     \frac{1}{2 \pi  \sigma _{T_b}  \sigma _{T_e}   \sqrt{1- \rho^2} h_b  h_e }  
    } \nonumber \\ \quad
		&\times & \exp \left[- \frac{1}{2 (1 - {\rho}^2)}   \left( 
 \frac{ \left( \textrm{ln}(h_b)  - \mu_{T_b} \right )^2}{  \sigma _{T_b}^2}   
- \frac{2 \rho     \left( \textrm{ln}(h_b)  - \mu_{T_b} \right )     \left( \textrm{ln}(h_e)  - \mu_{T_e} \right )    }{  \sigma _{T_b}   \sigma _{T_e}  }  
+ \frac{ \left( \textrm{ln}(h_e)  - \mu_{T_e} \right )^2}{  \sigma _{T_e}^2} 
\right ) \right ]      \label{eq_jointPDF}
\end{IEEEeqnarray}
	\setcounter{equation}{\value{tempequationcounter1}}
\end{figure*}

\section{Numerical Results and Discussions} \label{sec_dicussion}
Several statistical channel models have been proposed in the literature for the distribution of turbulence-induced fading in FSO systems. 
It is well accepted that for weak to moderate turbulence strength the log-normal model is valid,
 for moderate to strong turbulence the gamma-gamma model is effective, 
and in the limit of strong turbulence (i.e., in saturation regime and beyond) 
the negative exponential model is suitable \cite{Ghassemlooy2012Optical}.

\subsection{Log-normally distributed irradiance model}
We now specializes the secrecy capacity result to the log-normally distributed irradiance model, 
which applies to the FSO systems deployed over relatively short ranges in urban areas \cite{Khalighi2014Survey}.
Recall that $H_u $ is a mean-normalized irradiance fading RV. 
In the log-normal fading model, the PDF of $H_u $ is expressed as 
\begin{equation}  
f_{H_u} (h) = \frac{1}{ \sqrt{2 \pi  \sigma _{T_u} ^2} h}   \exp  \! \left( \! -  \frac{\left(  \ln (h) - \mu_{T_u}  \right) ^2}{2 \sigma _{T_u} ^2} \! \right ), \: \: \:   h > 0
\end{equation}
where $H_u = e^{T_u}$ with $T_u \sim  \mathcal{N} (\mu_{T_u}, \sigma _{T_u} ^2)$, 
and $\sigma _{T_u} ^2$ is the Rytov variance, a  parameter characterizing the strength of atmospheric turbulence. 
Note that the mean-normalized condition $\mathbb{E} [H_u] = 1$ requires the choice of $\mu_{T_u} = - \sigma _{T_u} ^2  /2$.
In general, $\sigma _{T_u} ^2$ is a function of the refractive structure index $C_n^2 (\cdot)$ and the horizontal distance $L$ traveled by the optical radiation. 
$C_n^2 (\cdot)$ typically ranges from $10^{-12} \textrm{ m}^{-2/3}$ for the strong turbulence to $10^{-17} \textrm{ m}^{-2/3}$ for the weak turbulence 
with a typical average value being $10^{-15} \textrm{ m}^{-2/3}$.
However, for horizontal optical propagation, as is the case in most terrestrial applications, the refractive structure index $C_n^2$ is constant and the 
Rytov variance is thus given by \cite{Ghassemlooy2012Optical}
\begin{equation}   \label{eq_RytovVariance}
	\sigma _{T_u} ^2 = 1.23 C_n^2 k^{7/6} L^{11/6},
\end{equation}
where 
$k$ is the wave number (i.e., $k = 2 \pi / \lambda $ with $\lambda $ the optical wavelength). 
Note that for horizontal communication, $L$ equals $d_u$ in Section \ref{sec_sysModel}. 
The weak, moderate, and strong intensity fluctuations are associated with $\sigma _{T_u} ^2  < 1$, $\sigma _{T_u} ^2 \approx  1$ and $\sigma _{T_u} ^2 >  1$, respectively \cite{Andrews2005Laser}.
In table \ref{table_parameters}, we list pairs of $\sigma _{T_u} ^2$ 
and $L$ for a typical wavelength $\lambda =1.5 \mathrm{\, \mu m}$ and $C_{n}^{2} = 10 ^{-15} \mathrm{\, m}^{-2/3}$ based on (\ref{eq_RytovVariance}).

$\addtocounter{equation}{1}$ 
When Eve's location is close to Bob, $H_b$ and $H_e$ are assumed to be correlated as 
the signals received by them go through a similar atmospheric environment \cite{Abaza2014Spatial, Wang2012Performance, Lopez-Martinez2015Physical}.
Hence, their joint PDF \cite{Wang2012Performance} is given in (\ref{eq_jointPDF}) at the bottom of this page.
In (\ref{eq_jointPDF}), $\rho$ denotes the correlation coefficient between $T_b$ and $T_e$.
Similarly, the correlation coefficient between $H_b$ and $H_e$ is denoted by $\rho_{H}$, 
and it is related to $\rho$ as follows \cite{Wang2012Performance}
\begin{equation}  
	 \rho    =    \frac{   \textrm{ln} \left(  \rho_{H}   \sqrt{\left(\exp(  \sigma _{T_b}^2  )  -   1 \right )  
     \left(\exp(  \sigma _{T_e}^2  )  -   1 \right )}  + 1 \right )  }{ \sigma _{T_b}  \sigma _{T_e}  } .
\end{equation}
Here, $\rho_{H} $ can be calculated based on the experimental data and its mathematical definition
\begin{equation}  
	 \rho_{H}   =   \frac{ \textrm{Cov} (H_b, H_e) }{  \sqrt{ \textrm{Var} (H_b) \textrm{Var} (H_e)}  } .
\end{equation}
By substituting (\ref{eq_jointPDF}) into (\ref{eq_avgSecCapacityLB}), we can derive the lower bound of secrecy capacity under the log-normally distributed irradiance model.

\subsection{Numerical analysis}
\begin{figure*}[ht]
  \normalsize
	\begin{minipage}{.481\textwidth}
	\centering
  \includegraphics[width= 0.8 \linewidth]{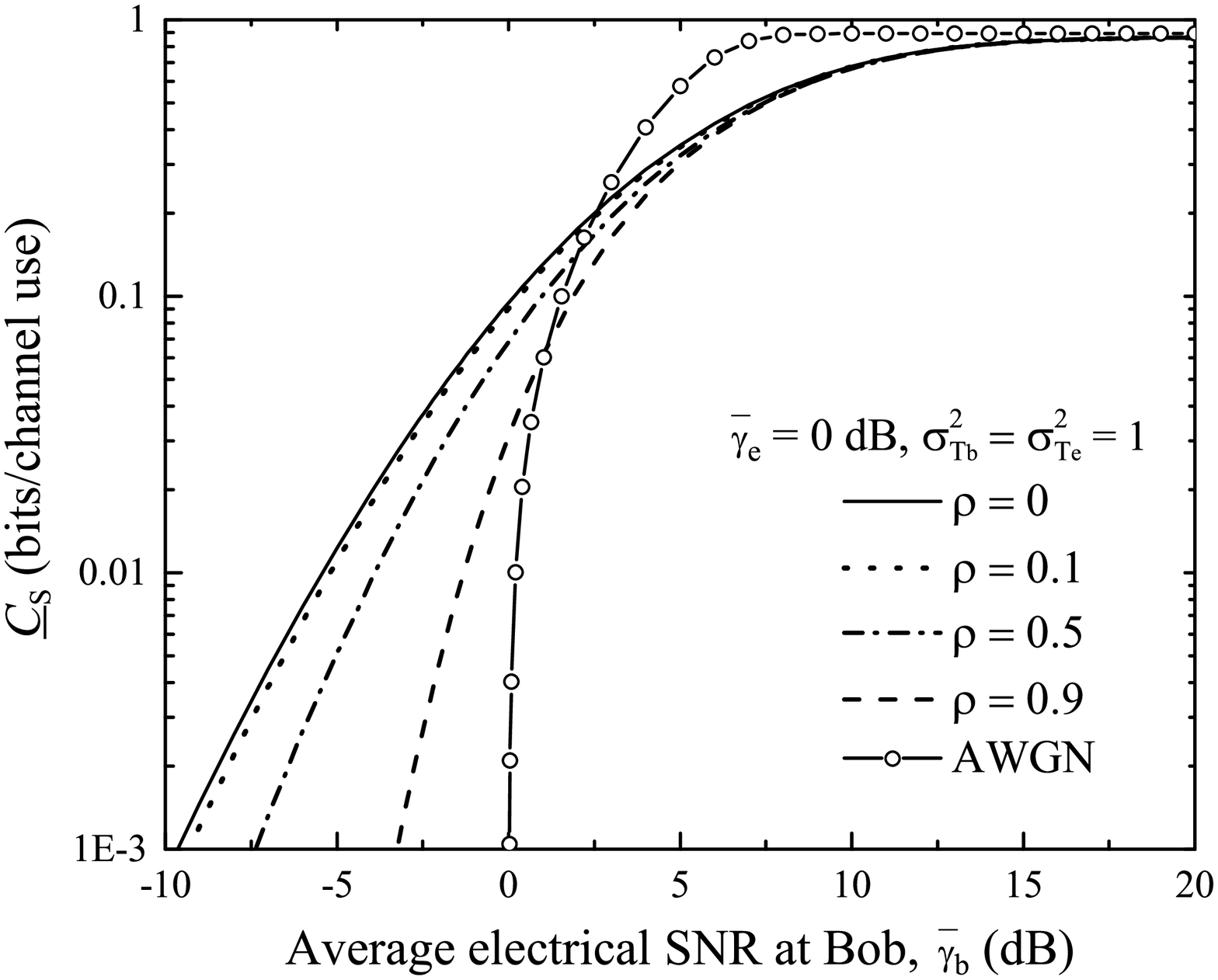}
\caption{$\underline{C_s}$ vs. $\overline{\gamma}_b$ under the correlated log-normal fading model with correlation coefficient $\rho = \{0, 0.1, 0.5, 0.9 \}$ and under the AWGN model.}
\label{fig:Cs_SNRbob_cor}
 \end{minipage}
\hfill
\begin{minipage}{.48\textwidth}
\centering
  \includegraphics[width= 0.8 \linewidth]{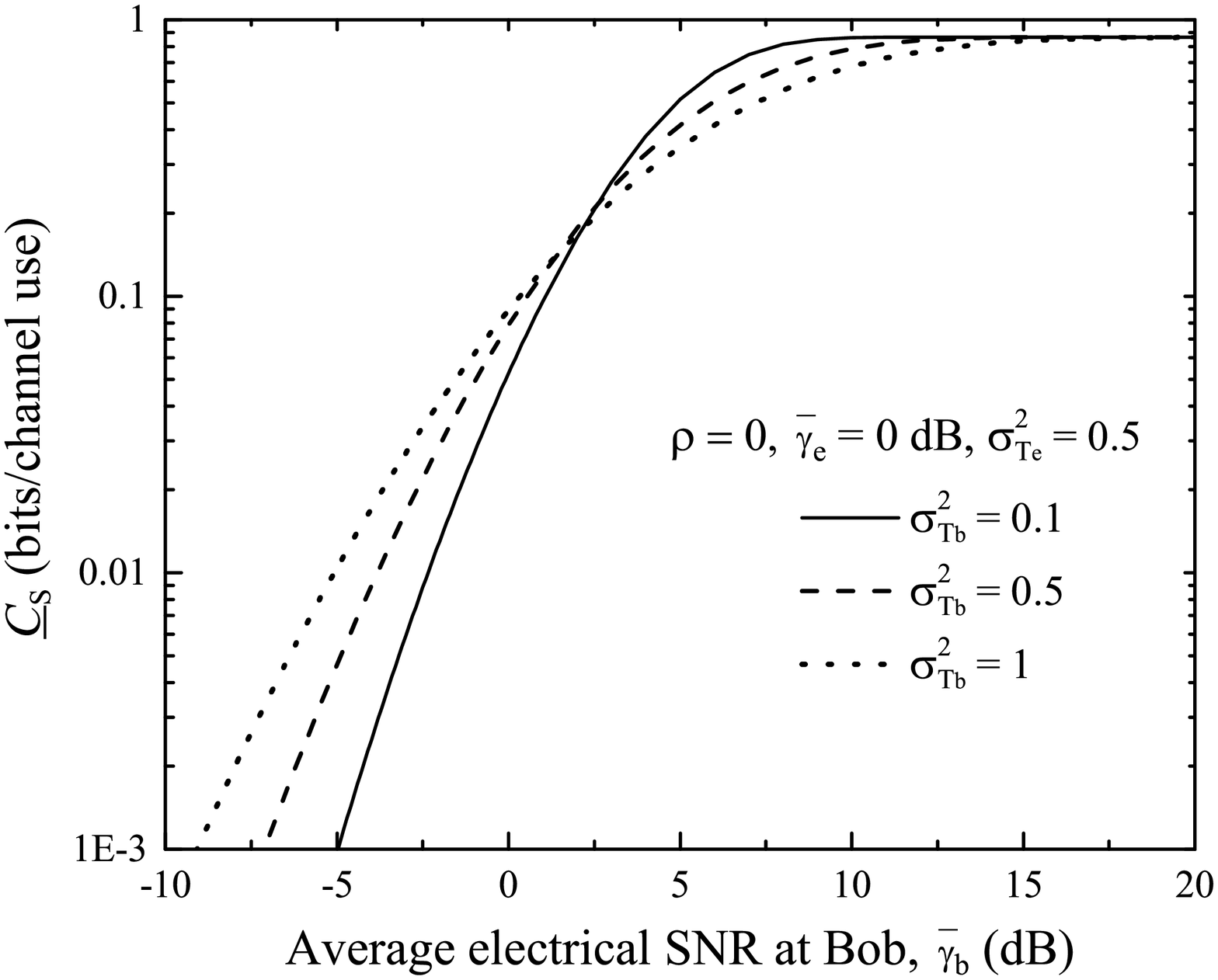}
\caption{$\underline{C_s}$ vs. $\overline{\gamma}_b$ for three cases of turbulence strength.}
\label{fig:Cs_SNRBob_var}
 \end{minipage}
\end{figure*}

Since even the numerical simulation of the exact average secrecy capacity is very difficult, we will focus on analyzing the numerical results of the lower bound of average secrecy capacity.
To illustrate the impact of correlation between Bob's and Eve's channels on the average secrecy capacity,
we show in Fig. \ref{fig:Cs_SNRbob_cor} how $\underline{C_s}$ varies with the \textbf{average electrical signal-to-noise ratio} (SNR) at Bob, i.e., $\overline{\gamma}_b=  \frac{S_b}{\sigma_ b}$, 
for a given average electrical SNR at Eve, i.e., $\overline{\gamma}_e = \frac{S_e}{\sigma_e}$,
under several channel correlation levels.
The parameters in Fig. \ref{fig:Cs_SNRbob_cor} are set to $\overline{\gamma}_e = 0 \textrm{ dB}$ and $\sigma _{T_b}^2 = \sigma _{T_e}^2 =1$.
Figure \ref{fig:Cs_SNRbob_cor} shows that
 for given $\overline{\gamma}_b$ and $\overline{\gamma}_e$, $\underline{C_s}$ is smaller when $H_b$ and $H_e$ are more correlated,
which indicates that the channel correlation has a negative impact on the average secrecy capacity.
However, such impact becomes very weak (almost disappeared) when $\overline{\gamma}_b \gg \overline{\gamma}_e$. 
Figure \ref{fig:Cs_SNRbob_cor} also shows that a moderate channel correlation (e.g., $\rho = 0.1$) does not have significant effect on the secrecy capacity.

\begin{table}[!t]
\renewcommand{\arraystretch}{1.6}
\caption{Parameters for $\lambda =1.5 \mathrm{\, \mu m}$ and $C_{n}^{2} = 10 ^{-15} \mathrm{\, m}^{-2/3}$ in weak atmospheric turbulence}
\vspace*{-4  pt}
\label{table_parameters}
\centering
\begin{tabular}{|l|l|l|l|}
\hline
$\sigma _{T_u}^2$ & $0.1$ & $0.5$ & $1$\\
\hline
$L $ & $2.4$ km & $5.7$  km & $8.3$  km \\
\hline
\end{tabular}
\vspace*{-8  pt}
\end{table}


To illustrate the impact of fading on the average secrecy capacity, we add 
in Fig. \ref{fig:Cs_SNRbob_cor} the similarly defined lower bound of secrecy capacity in additive white Gaussian (AWGN) channels  with the same $\overline{\gamma}_b$ and $\overline{\gamma}_e$ as in fading channels.
One can easily observe that, in AWGN channels, positive secrecy capacity cannot be achieved when $\overline{\gamma}_b \leq \overline{\gamma}_e$, i.e., $\overline{\gamma}_b \leq 0 \textrm{ dB}$ in Fig. \ref{fig:Cs_SNRbob_cor}.
On the contrary, in the presence of fading, positive secrecy capacity can still be achieved even when $\overline{\gamma}_b \leq \overline{\gamma}_e$. 
Note that a similar conclusion has been made in wireless RF communication \cite{Barros2006Secrecy}.
However, in the case $\overline{\gamma}_b$ is much larger than $\overline{\gamma}_e$, one can observe that $\underline{C_s}$ of the fading channels is smaller than that of the AWGN channels,
which indicates fading has a negative impact on the secrecy capacity when $\overline{\gamma}_b \gg \overline{\gamma}_e$.

 %

Figure \ref{fig:Cs_SNRBob_var} depicts how the turbulence (fading) strength actually affects the secrecy capacity 
in weak atmospheric turbulence for which the log-normal fading is valid.
We use three values of $\sigma _{T_b} ^2$ as given in Table \ref{table_parameters} while keeping $\sigma _{T_e} ^2$ fixed to show three different cases:  
$\sigma _{T_b} ^2 > \sigma _{T_e} ^2$, $\sigma _{T_b} ^2 = \sigma _{T_e} ^2$ and $\sigma _{T_b} ^2 < \sigma _{T_e} ^2$.
One can observe that in typical laser communication scenario with $\overline{\gamma}_b \gg \overline{\gamma}_e$ (e.g., $\overline{\gamma}_b = 10 \textrm{ dB}$ in Fig. \ref{fig:Cs_SNRBob_var}), 
$\underline{C_s}$ is larger for a weaker turbulence like $\sigma _{T_b} ^2 < \sigma _{T_e} ^2$, 
which indicates that the atmospheric turbulence has a negative impact on the secrecy capacity.
However, in other scenarios when $\overline{\gamma}_b$ is not that large (or maybe smaller) compared to $\overline{\gamma}_e$, the atmospheric turbulence actually is helpful to the secrecy capacity.

\section{Conclusion}  \label{sec_conclusion}
In this paper, we studied the problem of transmitting confidential information between two legitimate users over the FSO link in the presence of a malicious eavesdropper.
We have established a general formula for the average secrecy capacity of the laser-based FSO system with OOK modulation and threshold detection. 
We then specialized the formula to the particular correlated log-normal fading model and conducted numerical analysis.
Our results reveal that turbulence-induced fading $H_u$ as well as the turbulence strength $\sigma _{T_u} ^2$ has a negative impact on the secrecy capacity when Bob's channel condition is much better than Eve's (i.e., $\overline{\gamma}_b \gg \overline{\gamma}_e$),
but is possible to provide positive impact otherwise.
On the other hand, channel correlation is harmful to the secrecy capacity in all conditions, 
which agrees with the previous result regarding the impact of channel correlation on secrecy capacity in wireless RF communication \cite{Zhu2014ITC}.


%
%
%
%
%
%


%

\appendices
\section{Proof of Lemma \ref{lemma_optimumThresh}} \label{sec_proofLemma1}
When $q = 1/2$ (i.e., $P_X(1) = P_X(0) = 1/2$), we have
\begin{eqnarray}  \label{eq_capacityEve}
    \lefteqn{
      I(X;Y_e| q = \frac{1}{2})
    } \nonumber \\ \quad
    &=&  \mathcal{H}  \left(  \frac{1}{2} +  \frac{1}{2} \left( \varepsilon_0^e   - \varepsilon_1^e \right ) \right )	 
	- \frac{1}{2}   \left[ \mathcal{H}  \left(\varepsilon_0^e \right ) + \mathcal{H}  \left(\varepsilon_1^e \right ) \right], \  \  \  \  \  \    
\end{eqnarray}
where $\varepsilon_0^e  = \mathbb{P}_e (1|0)$ and $\varepsilon_1^e  = \mathbb{P}_e (0|1)$ are given in (\ref{eq_BERBob0_tmp}) and (\ref{eq_BERBob1_tmp}), respectively.
Replacing $\varepsilon_0^e = \frac{1}{2} \mathrm{erfc} \left( \frac{\tau _e}{ \sqrt{2 } \sigma_e} \right )$
and $\varepsilon_1^e = \frac{1}{2} \mathrm{erfc} \left( \frac{ S_{e} h_e - \tau _e}{ \sqrt{2 } \sigma_e} \right )$
into (\ref{eq_capacityEve}),
we then calculate the following maximum value 
\begin{equation}  \label{eq_capacityEve2}
\max_{\tau _e \in [0, \infty )}  \!\! \left\{  \mathcal{H} \!\!  \left(  \frac{1}{2} +  \frac{1}{2} \left( \varepsilon_0^e   - \varepsilon_1^e \right )  \!\! \right )	 
	- \frac{1}{2}   \left[ \mathcal{H}  \left(\varepsilon_0^e \right ) + \mathcal{H}  \left(\varepsilon_1^e \right ) \right]  \right\}.
\end{equation}
Let $f(\tau _e)  = I(X;Y_e| q = \frac{1}{2})$, and we expect that 
$f(\tau _e) $ is a concave function of $\tau _e$.
Denoting ${\tau _e}^*$ as the parameter achieving the maximum value of $\max_{\tau _e \in [0, \infty )}  f(\tau _e) $, 
we can derive ${\tau _e}^*$ by letting the first derivative of $f(\tau _e)$ with respect to $\tau _e$ equal to 0.
And we have
\begin{IEEEeqnarray}{rCl}
    \frac{\partial {f(\tau _e)}}{\partial \tau _e}    &=&      \frac{1}{2 \log 2}     \log \left(\frac{1 -  \left(\varepsilon_0^e   -  \varepsilon_1^e \right )}{ 1 +  \left(\varepsilon_0^e   -  \varepsilon_1^e \right ) }  \cdot \frac{ \varepsilon_0^e }{ 1-\varepsilon_0^e } \right )         \frac{\partial {\varepsilon_0^e}}{\partial \tau _e}    \nonumber \\
    && -   \frac{1}{2 \log 2}     \log \left(\frac{1 -  \left(\varepsilon_0^e   -  \varepsilon_1^e \right )}{ 1 +  \left(\varepsilon_0^e   -  \varepsilon_1^e \right ) }     \cdot    \frac{ 1-\varepsilon_1^e }{ \varepsilon_1^e } \right )         \frac{\partial {\varepsilon_1^e}}{\partial \tau _e}.  \  \  \    \  \  　 \label{eq_CapacityEve_der}
\end{IEEEeqnarray}
It can be easily validated that ${\tau _e}^* = S_{e} h_e/2$ makes $\frac{\partial {f(\tau _e)}}{\partial \tau _e} = 0$,
which indicates the maximum value is achieved at ${\tau _e}^* = S_{e} h_e/2$.   
Furthermore, numerical check has also been conducted to validate that ${\tau _e}^* = S_{e} h_e/2$ maximizes $f(\tau _e)$.
Therefore, Lemma \ref{lemma_optimumThresh} is proved.


%

\ifCLASSOPTIONcaptionsoff
  \newpage
\fi



\bibliographystyle{IEEEtran}
\bibliography{IEEEabrv,myReference}

%

%
%
%




\end{document}